\documentclass{sig-alternate}
\usepackage{epsf}
\usepackage{epsfig}
\usepackage{graphicx}
\usepackage{color}
\usepackage{mathrsfs}

\usepackage{latexsym}
\usepackage{amssymb}
\usepackage{amsmath}

\usepackage{stmaryrd}

\usepackage{verbatim}

\newcommand{\loss}{\mbox{\sc Loss}}
\newcommand{\gain}{\mbox{\sc Gain}}
\newcommand{\ema}{\mbox{\sc Ema}}
\newcommand{\rma}{\mbox{\sc Rma}}
\newcommand{\el}{m}
\newcommand{\rstar}{r}
\newcommand{\fstar}{\phi}
\newcommand{\G}{G}
\newcommand{\tstar}{{t(\phi,r)}}
\newcommand{\tfstar}{\phi}
\newcommand{\eff}{\mbox{\sc Eff}}

\newcommand{\common}{\Gamma}

\newtheorem{theorem}{Theorem}
\newtheorem{definition}{Definition}

\newtheorem{lemma}[theorem]{Lemma}

\newtheorem{fact}[theorem]{Fact}
\newtheorem{corollary}[theorem]{Corollary}

\newcounter{rem}
\begin{document}

\conferenceinfo{EC'09,} {July 6--10, 2009, Stanford, California,
USA.} \CopyrightYear{2009} \crdata{978-1-60558-458-4/09/07}

\title{Efficiency of (Revenue-)Optimal Mechanisms}
\date{}

\numberofauthors{3}
\author{
\alignauthor
Gagan Aggarwal\\
       \affaddr{Google Inc.}\\
       \affaddr{Mountain View, CA}\\
       \email{gagana@google.com}
\alignauthor Gagan Goel\titlenote{work done while the author was visiting Google Inc.}\\
       \affaddr{Georgia Tech}\\
       \affaddr{Atlanta, GA}\\
       \email{gagan@gatech.edu}
\alignauthor Aranyak Mehta\\
       \affaddr{Google Inc.}\\
       \affaddr{Mountain View, CA}\\
       \email{aranyak@google.com}
}

\maketitle
\begin{abstract}
We compare the expected efficiency of revenue maximizing (or {\em
optimal}) mechanisms with that of efficiency maximizing ones. We
show that the efficiency of the revenue maximizing mechanism for
selling a single item with $(k + \log_{\frac{e}{e-1}}{k} + 1)$
bidders is at least as much as the efficiency of the efficiency
maximizing mechanism with $k$ bidders, when bidder valuations are
drawn i.i.d. from a Monotone Hazard Rate distribution. Surprisingly,
we also show that this bound is tight within a small additive
constant of $5.7$. In other words, $\Theta(\log{k})$ extra bidders
suffice for the revenue maximizing mechanism to match the efficiency
of the efficiency maximizing mechanism, while $o(\log{k})$ do not.
This is in contrast to the result of Bulow and
Klemperer~\cite{bulow-klemperer} comparing the revenue of the two
mechanisms, where only one extra bidder suffices. More precisely,
they show that the revenue of the efficiency maximizing mechanism
with $k+1$ bidders is no less than the revenue of the revenue
maximizing mechanism with $k$ bidders.

We extend our result for the case of selling $t$ identical items and
show that $2.2\log{k} + t \Theta(\log{\log{k}})$ extra bidders
suffice for the revenue maximizing mechanism to match the efficiency
of the efficiency maximizing mechanism.

In order to prove our results, we do a classification of
Monotone Hazard Rate (MHR) distributions and identify a family of MHR
distributions, such that for each class in our classification, there
is a member of this family that is pointwise lower than every
distribution in that class. This lets us prove interesting structural
theorems about distributions with Monotone Hazard Rate.

\end{abstract}




\section{Introduction}

Auctions are bid-based mechanisms for buying and selling of goods.
The two most common objective functions in auction design are
efficiency and revenue. Efficiency is the sum of the surplus of both
the seller and the buyer, which represents the total social welfare,
whereas revenue is the surplus of the seller only. The efficiency
and revenue of auctions has been the subject of extensive study in
auction theory (see, e.g., the survey \cite{zhan-survey}, and the
citations within).
Ideally we would like to simultaneously maximize both the objective
functions. But these two goals cannot be achieved simultaneously.
Thus, we have the celebrated Vickrey-Clarke-Groves (VCG) mechanism
\cite{vickrey,clarke,groves} which is a truthful mechanism that
maximizes efficiency on the one hand, and Myerson's Optimal
Mechanism~\cite{myerson} that maximizes revenue on the other. If one
is interested in both objective functions, then one has to trade-off
the two. The exact balance of how much weight to give to each
objective is, perhaps, a difficult question in any real-world
setting. Generally, private sellers would like to maximize revenue,
but keeping long-term benefit of the business in mind, they might want
to keep social welfare high as well. Similarly, the primary goal in
allocating public goods is to maximize social welfare, but a secondary
objective might be to raise revenue.

In the light of this dilemma, a natural question that arises is: how
sub-optimal is the revenue of the VCG mechanism, and how sub-optimal
is the efficiency of Myerson's Optimal Mechanism. Bulow and
Klemperer \cite{bulow-klemperer} give a structural theorem which
characterizes the sub-optimality of the revenue of the VCG
mechanism. They show that for the case of selling a single item, the
VCG mechanism with one extra bidder makes at least as much revenue
in expectation as the expected revenue of Myerson's Optimal
mechanism (when bidder valuations are drawn from a class of
distributions called regular distributions).
A seller, who is currently using the VCG mechanism and wishes to
increase her revenue, faces two choices: (1) increase the reserve
price closer to Myerson's reserve price, or (2) attract more bidders
by investing in sales and marketing. Bulow and Klemperer's theorem
gives an insight into the trade-offs between these two choices.

In this paper, we characterize the sub-optimality of the efficiency of
Myerson's Optimal mechanism.
We show that, surprisingly, there exists a class of distributions with
monotone hazard rate for which a constant number of extra bidders does
not suffice for Myerson's optimal mechanism to match the efficiency of
the VCG mechanism. In fact, we show that one needs at least
$\omega(\log{k})$ extra bidders, where $k$ is the number of bidders
participating in the VCG mechanism. We match this lower bound by
showing that, for all monotone hazard rate distributions, $O(\log{k})$
extra bidders always suffice (our upper and lower bounds are tight up
to a small additive constant). This contradicts the following
intuition: since the efficiency of Myerson's Optimal mechanism gets
closer to the VCG mechanism as the number of bidders $k$ increases, we
would expect the number of extra bidders needed would go down as $k$
increases.  Another way of interpreting our upper bound is that with
$O(\log{k})$ extra bidders, Myerson's Optimal mechanism simultaneously
maximizes both revenue and efficiency. We extend our upper bound
result to the case of selling multiple identical items as well.

In order to prove the above results, we do a classification of
Monotone Hazard Rate (MHR) distributions and identify a family of MHR
distributions, such that for each class in our classification, there
is a member of this family that is pointwise lower than every
distribution in that class. This enables us to prove certain important
structural properties of distributions with Monotone Hazard Rate,
which helps us prove our main theorems.

\subsection{Model}

Our model consists of a single seller and $k$ buyers (bidders). We
will consider the case of selling a single item, and also the case of
selling $t$ identical items when each bidder has unit demand. The
private values $(v_i)_{i\in[k]}$ of the bidders are drawn
independently from a common distribution $D$. Here $v_i$ represents
the value of the bidder $i$ for one unit of the item. We will use
$f_D$ and $F_D$ to denote the probability density function and
cumulative distribution function of the distribution $D$ respectively.

The {\em hazard rate} of a distribution $D$ is given by $h_D(x)$ $:=
f_D(x)/(1-F_D(x))$. A distribution $D$ is said to have a {\em
monotone hazard rate} (MHR) if $h_D(.)$ is a non-decreasing function
of $x$. For the most part, we will assume (as is common in the
economics literature) that the given distribution $D$ has a monotone
hazard rate. Many common families of distributions such as the
Uniform and the Exponential families have MHR. We will assume that
the support of $D$ lies between $[0,\infty)$. Also, for the ease of
presenting main ideas, we will assume that $F_D$ is a continuous
function even though the results hold in non-continuous case as
well.

We will restrict our attention to truthful auctions, those in which
bidders have no incentive to misreport their true valuation. Thus we
can assume that the bidders bid their true private values, i.e., the bid
vector $(b_i)_{i\in[k]}$ is same as the value vector
$(v_i)_{i\in[k]}$. From here onwards, we will use the terms {\em bid}
and {\em value} interchangeably.

For a given mechanism, its {\em efficiency} on a given input is
defined as the sum of the valuations of the bidders who get the
good, while its {\em revenue} is defined as sum of the payments to
the seller. Since the private values of bidders are not arbitrary
but rather drawn from a distribution, we will be interested in the
values of efficiency and revenue {\em in expectation}. We will use
$\eff(M)$ to denote the expected efficiency of a mechanism $M$. \\

\subsection{Optimal Auctions: Vickrey and Myerson}
We will use $\ema(k)$ to denote the efficiency maximizing (VCG)
auction with k bidders ($D$ will always be clear from context). It
assigns the item to the highest bidder and charges it the second
highest bid. In the case of selling $t$ identical items, $\ema(k)$
allocates the items to the $t$ highest bidders, charging each of
them the $t+1$th highest bid.

Similarly, we will use $\rma(k)$ to denote revenue maximizing
auction (Myerson's auction) with $k$ bidders.
Myerson \cite{myerson} defined a notion of {\em virtual valuation}
$\psi_i$ of a bidder $i$, where $$\psi_i := v_i - \frac{1}{h_D(v_i)}$$
Myerson showed that the revenue maximizing truthful auction is the one
which maximizes the virtual efficiency (sum of the virtual valuations
of the auction winners). Thus, in the single item case, it assigns the
item to the bidder with the highest non-negative virtual value (and
does not sell the item if all virtual values are negative). If a
distribution $D$ satisfies the {\em regularity} condition, defined as
$\psi(x)$ being a non-decreasing function, then the above condition is
equivalent to assigning item to the highest bidder as long his virtual
value is non-negative (note that distributions with MHR always satisfy
the regularity condition). This cutoff value at which $\psi(x)=0$ is
called the reserve price $\rstar_D$.

\begin{definition}[Reserve price]
Reserve price of distribution $D$ is defined as:
$$r_D :=~ x, ~~ s.t.~~ h_D(x) = 1/x$$\
\end{definition}
We drop the subscript $D$, if $D$ is clear from context. Thus in the
single item case with regular distributions, $\rma(k)$ assigns the
item to the highest bidder, as long as its bid is no smaller than the
reserve price, and charges it the maximum of the reserve price and the
second highest bid (it does not sell the item if all bids are below
the reserve price). In the case of selling $t$ identical items,
$\rma(k)$ finds the $k'\leq k$ bidders whose bids are above the
reserve price, allocates one item each to the highest $\min\{t, k'\}$
bidders and charges each one the maximum of the reserve price and the
$t+1$th highest bid.

\subsection{Related Work}
Bulow and Klemperer \cite{bulow-klemperer} characterized the revenue
sub-optimality of $\ema$. They showed that $\ema(k+1)$ (with one
extra bidder) has at least as much expected revenue as $\rma(k)$.
Their result can be interpreted in a bi-criteria sense; VCG auctions
with one extra bidder simultaneously maximize both revenue and
efficiency. For the case of $t$ identical items, they show that $t$
additional bidders are needed for the result to hold.

In \cite{roughgarden-sundararajan}, Roughgarden and Sundararajan gave the {\em
  approximation factor} of the optimal revenue that is obtained by
$\ema(k)$. They show that, for $t$ identical items and $k$ bidders
with unit demand, the revenue of $\ema(k)$ is at least $(1 - t/k)$
times the revenue of $\rma(k)$. Neeman~\cite{Neeman} also studied the
percentage of revenue which $\ema(k)$ makes compared to $\rma(k)$ in
the single item case. \cite{Neeman} used a numerical analysis approach
and assumed that the distribution $D$ is any general distribution (not
restricted to regular or MHR as in \cite{roughgarden-sundararajan}) but with a
bounded support.

In another related work looking at simultaneously optimization of both
revenue and efficiency, Likhodedov and Sandholm \cite{likhodedov-sandholm} gave a
mechanism which maximizes efficiency, given a lower bound constraint
on the total revenue.

\subsection{Our Results}

We study the number of extra bidders required for Myerson's
(revenue-)optimal mechanism to achieve at least as much efficiency
as the efficiency maximizing (VCG) mechanism.  Let $\alpha = 1-
1/e$. For bidder valuations drawn from a distribution with Monotone
Hazard Rate, we prove the following:

\begin{itemize}
\item {\em Single item Upper Bound (Theorem~\ref{thm:1item-ub}):} In
the single item case, $\el \geq
\left\lfloor\log_{\frac{1}{\alpha}}{2k} \right\rfloor + 2$ extra
bidders suffice to for the revenue maximizing mechanism to achieve
at least as much efficiency as the efficiency maximizing mechanism
with $k$ bidders (for any $k$).

\item {\em Single item Lower Bound (Theorem~\ref{thm:1item-lb}):} In
the single item case, we demonstrate a distribution having monotone
hazard rate, such that for any $k$, if the efficiency-optimal
mechanism has $k$ bidders, and the revenue-optimal mechanism has \\
$\left\lfloor\log_{1/\alpha}{(k+1)(1-\alpha)}\right\rfloor + 1$
extra bidders, then the efficiency of the latter is strictly less
than the efficiency of the former.  In other words, \\
$\left\lfloor\log_{1/\alpha}{(k+1)(1-\alpha)}\right\rfloor + 1$
extra bidders do not suffice.

\item {\em Multi item Upper Bound (Theorem~\ref{thm:multiitem-ub}):} In
the case of selling $t$ identical items, with bidders having
unit-demand: $\el+s$ extra bidders suffice for the revenue-optimal
mechanism to achieve at least as much efficiency as the
efficiency-optimal mechanism with $k$ bidders, where $\el =
\left\lfloor\log_{\frac{1}{\alpha}}{2k} \right\rfloor + 2$ and $s
\geq t + (1+\epsilon)t\log{\el}$, for every $\epsilon
>0$ and large enough $k$. Thus,
approximately, $\log{k} + t \log\log{k}$ extra bidders suffice.

\item We also show that if both auctions have the same number of
bidders $k$, then the ratio of the efficiency of the revenue-optimal
auction to the optimal efficiency is at least $1 - \alpha^k$. (The
proof is easy and we omit it in this extended abstract\footnote{In
fact we can also prove using the same techniques that the ratio of the
revenue of VCG to that of Myerson's Optimal Auction is
$1-\alpha^{k-1}$, which improves on the polynomial bound provided
in~\cite{roughgarden-sundararajan}. }).

\item We also prove (Section~\ref{sec:regular}) that our upper bound
result does not hold for regular distributions -- we show that for
every $k, m$, there is a regular distribution for which the efficiency
of the revenue-optimal mechanism with $k+m$ bidders is strictly lower
than the efficiency of the efficiency-optimal mechanism with $k$
bidders.

\end{itemize}

\subsection*{Outline of the paper:}
Section~\ref{sec:setup} describes some basic setup which is common to
the rest of the paper. Section~\ref{sec:1item} describes our results
for the single item case. We start with the definition of two
quantities $\gain$ and $\loss$, and analyze the expression $\gain -
\loss$ in
Section~\ref{subsec:gain-loss}. Section~\ref{subsec:1-item-ub} and
~\ref{subsec:1-item-lb} prove our upper and lower bound results
respectively. In Section~\ref{sec:multiitem}, we extend out upper
bound result to the case of selling $t$ indentical items to bidders
with unit-demand. Section~\ref{sec:regular} deals with the case of
regular distributions.

\section{Basic Setup}
\label{sec:setup}

%
%
%
%
%
%
%
%
%
%

To compare the two mechanisms $\rma(k+\el)$ and $\ema(k)$, which have
a different number of bidders, we will think of the process of drawing
their bidder values as drawing $k+\el$ bidder values
$(b_1,b_2,..b_{k+\el})$ independently from the distribution $D$ -- the first k
bidders ($b_1, b_2,..b_k$) participate in both $\rma(k+\el)$ and
$\ema(k)$, whereas the last $\el$ bidders participate only in
$\rma(k+\el)$.

The following lemma will prove useful in the subsequent sections.
\begin{lemma}\label{max.F}
For any MHR distribution D: $F_D(\rstar) \leq 1-1/e$, where $\rstar$ is the
reserve price for the distribution $D$.
\end{lemma}
\begin{proof}
Let the hazard rate of distribution D be $h(x)$. By definition of
hazard rate, we have $F_D(x) = 1- e^{-\int_0^x h(t) dt}$. By
definition of reserve price, $h(\rstar)= 1/\rstar$. Also since D has
MHR, we get $h(x) \leq 1/\rstar$, $\forall x \leq \rstar$. Thus,
\begin{align}
\forall x \leq \rstar: ~~~1-e^{-\int_0^x h(t) dt}~&\leq~1-e^{-\int_0^x 1/\rstar
dt} \notag\\
\Rightarrow F_D(\rstar) ~ &\leq~1-e^{-\int_0^\rstar 1/\rstar dt} \notag \\
 &~~~~~~~~(\mbox{setting } x = \rstar)\notag \\
\Rightarrow F_D(\rstar)~&\leq ~ 1-e^{-1}\notag
\end{align}
\end{proof}

\section{Selling one item}
\label{sec:1item}

We begin by noting that when selling a single item, if the value of
any of the first $k$ bidders is greater than or equal to the reserve
price $\rstar_D$, then $\rma(k+\el)$ achieves at least as much
efficiency as $\ema(k)$.

The more challenging case (for the upper bound) occurs when the value
drawn by all the first $k$ bidders is less than the reserve price. In
this case, $\ema(k)$ achieves an efficiency equal to the highest value
among the values of the first $k$ bidders, whereas the contribution of
the first $k$ bidders to $\rma(k+\el)$ is zero as all of these bidders
have a value less than the reserve price cutoff. In other words,
conditioned on the event that first $k$ bidders' value is less than
the reserve price $\rstar_D$, the expected efficiency of $\ema(k)$ is
$\frac{\int_0^{\rstar} x f^{(k)}(x) dx}{F^{(k)}(\rstar)}$, where
$F_D^{(k)}$ and $f_D^{(k)}$ are the c.d.f and p.d.f of the maximum of
$k$ numbers picked i.i.d. from $D$ (we will drop the subscript $D$
whenever $D$ is clear from the context). Note that $F^{(k)}(x) =
F^k(x)$, and $f^{(k)}(x) = kF^{k-1}(x)f(x)$. Also, conditioned on this
event, the expected contribution of the first $k$ bidders
$(b_1,b_2,...,b_k)$ to the efficiency of $\rma(k+l)$ is zero. We
define:
\begin{equation}
\loss_D = \frac{\int_0^{\rstar} x f_D^{(k)}(x) dx}{F_D^{k}(\rstar)}
\label{eqn:loss}
\end{equation}

To make up for this lost efficiency, the revenue maximizing
mechanism has $\el$ extra bidders. The contribution to the expected
efficiency of $\rma(k+\el)$ from these $\el$ extra bidders
$(b_{k+1}, b_{k+2},.. b_{k+\el})$, conditioned on the event that the
value of bidders $(b_1, b_2, ..b_k)$ is less than $\rstar_D$, is at
least $(1 - F^{\el}(\rstar_D))\rstar_D$. This is because of the fact
that all the draws are independent and the probability that at least
one of the $\el$ extra bidders will have a value higher than the
reserve price $\rstar_D$ is $(1 - F^{\el}(\rstar_D))$. We define:
\begin{equation}
\gain_D = (1 -F^{\el}(\rstar_D))\rstar_D
\label{eqn:gain}
\end{equation}

If for all distributions $D$, it is true that $\gain_D - \loss_D \geq
0$ for some $k, \el$, then we know that the efficiency of
$\rma(k+\el)$ is at least as much as the efficiency of
$\ema(k)$. Moreover, if we can demonstrate a distribution $D$ s.t. the
expected contribution to the efficiency of $\rma(k+\el)$ from these
extra $\el$ bidders $(b_{k+1}, b_{k+2},.. b_{k+\el})$, is strictly
less than $\gain_D$, i.e., if we show that $\gain_D - \loss_D < 0 $,
for some $k, l$, (and that there is no additional gain to
$\rma(k+\el)$ in the case when one of the first $k$ bidders has value
equal to or greater than $\rstar_D$), then we would have shown that
$\el$ extra bidders does not suffice.

Therefore, we examine this key expression $\gain_D - \loss_D$ in the next section.  We will omit the
subscript $D$ whenever it is clear from the context.

\subsection{The expression $\gain - \loss$}
\label{subsec:gain-loss}

Recall the expressions $\gain_D$ and $\loss_D$ as defined in
Equations~\ref{eqn:gain} and~\ref{eqn:loss}.

For the purpose of getting a better handle on the expression $\gain -
\loss$, we will partition the set of all MHR distributions into
different classes, according to their optimal reserve price $\rstar$ and the
value of the cdf at the optimal reserve price, as follows. Let
$\mathcal{D}(\rstar,\fstar)$ be the set of MHR distributions with a
fixed reserve price $\rstar \geq 0$ and $F(\rstar)= \fstar$. Note
that, by Lemma~\ref{max.F}, $\mathcal{D}(\rstar, \fstar)$ is non-empty
only if $\fstar \in [0,1-1/e]$. Also note that all these distributions
have the same value for the expression $\gain$ and differ only in the
numerator of the expression $\loss$.

Next, we find a distribution in $\mathcal{D}(\rstar,\fstar)$ which
maximizes the numerator of $\loss$.
\begin{definition}[Distribution ${\G}_{\tfstar,\rstar}$]
\label{def:Gfr} Let $~\rstar \geq 0$, $\fstar \in [0,1-1/e]$, and
let $~\tstar = \rstar (1+ \ln(1-\fstar)) \in [0,r]$. Then,
\begin{equation*}
  {\G}_{\tfstar,\rstar}(x)=
 \begin{cases} 0 & \text{~~~~$0\leq x < \tstar$,}
    \\
     1 ~-~ e^{-\frac{1}{\rstar}(x-\tstar)} &\text{~~~~$\tstar \leq x \leq \rstar$,}
     \\
     \fstar + \frac{1-\phi}{\epsilon}.(x - \rstar)  &\text{~~~~$\rstar \leq x \leq r + \epsilon$,}
     \\
     1 &\text{~~~~$x \geq r+\epsilon.$}
  \end{cases}\notag
\end{equation*}
\end{definition}

Here $\epsilon$ is any positive number. It can be verified that the
above distribution has the following properties:
\begin{itemize}
\item ${\G}_{\tfstar,\rstar}$ is MHR.
\item The optimal reserve price for ${\G}_{\tfstar,\rstar}$ is
  $\rstar$.
\item ${\G}_{\tfstar,\rstar}(\rstar) = \fstar$. Therefore,
${\G}_{\tfstar,\rstar} \in \mathcal{D}(\rstar,\fstar)$.
\end{itemize}

Next, we prove that ${\G}_{\tfstar, \rstar}$ is (point-wise) no larger than
every other function in $\mathcal{D}(\rstar,\fstar)$

\begin{lemma}
\label{lem:F.P} For every distribution $D \in
\mathcal{D}(\rstar,\fstar)$, and any $y \in [0,\rstar]$:
$$F_D(y) \geq {\G}_{\tfstar,\rstar}(y)$$
\end{lemma}
\begin{proof}
The cdf of a distribution $D$ with hazard rate $h_D(.)$ can be written
as $F_D(y) = 1 - e^{-\int_0^y h_D(z) dz}$.

Now, by definition, $\rstar - \frac{1}{h_D(\rstar)}= 0$, implying $h_D(\rstar) =
1/\rstar$. Since $h_D(.)$ is an increasing function, therefore $h_D(z)
\leq 1/\rstar$ for all $z \leq \rstar$. Also note that $h_{{\G}_{\tfstar,\rstar}}(z) = 0$
for $z < \tstar$, and $h_{{\G}_{\tfstar,\rstar}}(z) = 1/\rstar$ for $z \in [\tstar,\rstar]$.

Thus, for any $y\in[\tstar,\rstar]$ we have

\begin{align*}
&  \int_y^{\rstar} h_D(z)dz  \leq \int_y^{\rstar}
  h_{{\G}_{\tfstar,\rstar}}(z) dz \\
\Rightarrow~~ & \int_0^{\rstar} h_D(z)dz - \int_0^{y} h_D(z)dz\notag
\\
 & \leq \int_0^{\rstar} h_{{\G}_{\tfstar,\rstar}}(z) dz - \int_0^{y}
  h_{{\G}_{\tfstar,\rstar}}(z) dz
\end{align*}

Since $F_D(\rstar) = {\G}_{\tfstar,\rstar}(\rstar) = \fstar$, (and
recalling from the definition of hazard rate that $F_D(x) = 1 -
e^{-\int_0^x h_D(t) dt}$ for all distributions $D$) we have
$\int_0^{\rstar} h_D(z)dz =$ \\ $\int_0^{\rstar}
h_{{\G}_{\tfstar,\rstar}}(z) dz$.  Therefore,

$$ - \int_0^{y} h_D(z)dz \leq - \int_0^{y} h_{{\G}_{\tfstar,\rstar}}(z) dz $$

$$\Rightarrow~~ 1 - e^{- \int_0^{y} h_D(z)dz} \geq 1 - e^{- \int_0^{y}
h_{{\G}_{\tfstar,\rstar}}(z) dz}$$

Thus, for any $y \in [\tstar,\rstar]$, $F_D(y)\geq {\G}_{\tfstar,\rstar}(y)$. Since
${\G}_{\tfstar,\rstar}(y)=0$ for $y\in[0,\tstar)$, we have proved the lemma.
\end{proof}

\vspace{1ex}

Next we prove that the numerator of $\loss$ among $\mathcal{D}(\rstar,\fstar)$ is
maximized at ${\G}_{\tfstar,\rstar}$.

\begin{lemma}
\label{cor.exp}
For every distribution $D \in \mathcal{D}(\rstar,\fstar)$: $$\int_0^{\rstar} x
f^{(k)}_D(x) dx \leq \int_0^{\rstar} x [{\G}^{(k)}_{\tfstar,\rstar}(x)]' dx$$
\end{lemma}
\begin{proof}
For any $D \in \mathcal{D}(\rstar,\fstar)$, we have
\begin{align}
\int_0^{\rstar} x f^{(k)}_D(x) dx & = \rstar F^{(k)}_D(\rstar) - \int_0^{\rstar}
F^{(k)}_D(x) dx \notag \\
&~~~~~ \mbox{(integrating by parts)} \notag \\
& \leq  \rstar {\G}_{\tfstar,\rstar}^{(k)}(\rstar) - \int_0^{\rstar} {\G}_{\tfstar,\rstar}^{(k)}(x) dx
\notag \\
& = \int_0^{\rstar} x [{\G}^{(k)}_{\tfstar,\rstar}(x)]' dx \notag
\end{align}

The inequality follows from Lemma~\ref{lem:F.P} and the fact that
$\rstar F^{(k)}_D(\rstar) =
\rstar {\G}_{\tfstar,\rstar}^{(k)}(\rstar) = \rstar \fstar^k$.
\end{proof}

\vspace{1ex}

Now that we know that, among all distributions in $D(\rstar, \fstar)$,
${\G}_{\tfstar,\rstar}$ maximizes the expression $\loss$, we can
calculate the maximum value of $\loss$ in terms of $\rstar$ and
$\fstar$. Lemma~\ref{lem:exp} examines the numerator of $\loss$ at
${\G}_{\tfstar,\rstar}$.

\begin{fact}
For a fixed $\lambda,~t$, and $k$: $ \int (1 - e^{-\lambda(x-t)})^k dx  = x - \frac{1}{\lambda}  \sum_{i=1}^k \frac{(1 -
e^{-\lambda(x-t)})^i}{i}$
\end{fact}

\begin{proof}
Let $g_k = \int (1 - e^{-\lambda(x-t)})^k dx$. We have
\begin{align*}
&g_k = \int (1 - e^{-\lambda(x-t)})^{k-1} dx  \\
&~~~~~- \int e^{-\lambda(x-t)}(1 - e^{-\lambda(x-t)})^{k-1}dx\\
\Rightarrow ~~& g_k = g_{k-1} - \frac{1}{\lambda .k} (1 -
e^{-\lambda(x-t)})^k \\
\Rightarrow ~~& g_k = g_0 - \frac{1}{\lambda}  \sum_{i=1}^k \frac{(1 -
  e^{-\lambda(x-t)})^i}{i} \\
\Rightarrow ~~& g_k = x - \frac{1}{\lambda}  \sum_{i=1}^k \frac{(1 - e^{-\lambda(x-t)})^i}{i}
\end{align*}
\end{proof}

\vspace{1ex}

\begin{corollary}\label{exp.int}
For a fixed $r$, $t$ and $k$, \\$ \int_{t}^{\rstar} (1 -
e^{-\frac{x-t}{\rstar}})^k dx   =  \rstar - t -  \rstar \sum_{i=1}^k
\frac{\fstar^i}{i}$, where \\ $\fstar = 1 -
e^{-\frac{\rstar-t}{\rstar}}$.
\end{corollary}

\vspace{1ex}

\begin{lemma}
\label{lem:exp}
 $\int_0^{\rstar} x [{\G}^{(k)}_{\tfstar,\rstar}(x)]' dx \\ ~~~~~~~~~~~~~~~= \rstar \left(\fstar^k ~+~ \ln(1-\fstar) + \sum_{i=1}^k \frac{\fstar^i}{i} \right)$
\end{lemma}
\begin{proof}
Integrating by parts, we have
\begin{align*}
\int_0^{\rstar} x [{\G}^{(k)}_{\tfstar,\rstar}(x)]'dx & = \rstar
\fstar^k - \int_0^{\rstar} {\G}^{(k)}_{\tfstar,\rstar}(x) dx \\
& =  \rstar \fstar^k -  \int_{\tstar}^{\rstar} (1 ~-~
e^{-\frac{1}{\rstar}(x-\tstar)})^k dx \\
& =  \rstar \fstar^k -  \left[ \rstar - \tstar - \rstar\sum_{i=1}^k
  \frac{\fstar^i}{i} \right] \\
&~~~~~   \mbox{~~~~~~~(by Corollary \ref{exp.int})}\\
&~~= \rstar \fstar^k + \rstar( 1+ \ln(1-\fstar)) - \rstar + \rstar\sum_{i=1}^k \frac{\fstar^i}{i}\\
&~~~~~~~~~~(\mbox{by definition of }\tstar)  \\
&~~= \rstar\left(\fstar^k + \ln(1-\fstar) + \sum_{i=1}^k \frac{\fstar^i}{i} \right)
\end{align*}
\end{proof}

\vspace{1ex}

Let $D \in \mathcal{D}(\rstar,\fstar)$. Then, for any given $\el$, from
equations~\ref{eqn:gain} and~\ref{eqn:loss} and Lemma~\ref{lem:exp},
we have:
\begin{align}
&\gain_D - \loss_D \notag\\
&~~ \geq  \gain_{{\G}_{\tfstar,\rstar}} - \loss_{{\G}_{\tfstar,\rstar} } \notag \\
&~~ =  (1-\fstar^{\el})\rstar - \frac{\rstar\left(\fstar^k +
  \ln(1-\fstar)  + ~\sum_{i=1}^k \frac{\fstar^i}{i}\right)}{\fstar^k} \notag \\
&~~ = \frac{\rstar\left(- (\fstar)^{k+\el} - \ln(1-\fstar) -
~\sum_{i=1}^k
      \frac{\fstar^i}{i}   \right)}{\fstar^k} \label{eqn:gain-loss-final}
\end{align}

\subsection{Upper bound on the number of extra bidders required}
\label{subsec:1-item-ub}
To prove an upper bound for distributions in $\mathcal{D}(\rstar, \fstar)$, we need to
find values of $\el$ (as a function of $k$) for which the expression
in Equation~\ref{eqn:gain-loss-final} is non-negative. To do this, we define a
uni-variate polynomial $$q(x) := x^{k+\el}+ \ln(1-x) + \sum_{i=1}^k
\frac{x^i}{i}$$ Since $\fstar \in [0, 1-1/e]$ by Lemma~\ref{max.F}, it
suffices to find $\el$ for which $q(x) \leq 0$ for all $x \in [0,
1-1/e]$.

Since $q(0) = 0$, if we find values of $\el$ for which $q'(x) \leq
0,~\forall x \in [0, 1-1/e]$, then $q(x) \leq 0~\forall x \in [0,
1-1/e]$. Now,
\begin{align}
q'(x) &= (k+\el)x^{k+\el-1} - \frac{1}{1-x}  + \sum_{i=1}^k x^{i-1}\notag\\
&= (k+\el)x^{k+\el-1} - \frac{1}{1-x} + \frac{1 - x^k}{1-x}\notag\\
&= (k+\el)x^{k+\el-1} - \frac{x^k}{1-x} \notag\\
&= \frac{x^k}{1-x} ( (k+\el)x^{\el-1}(1-x) - 1 )\notag
\end{align}

Since $\frac{x^k}{1-x} \geq 0$, $q'(x) \leq 0$ iff $x^{\el-1}(1-x) \leq
\frac{1}{k+\el}$. It is easy to see that, for $x\in[0,1-1/e]$,
$x^{\el-1}(1-x)$ is maximized at $x= 1-1/e$ for $\el > 2$.
Let $c =
\frac{e}{e-1}$, and $\el = log_c(2k) + 2$. For this choice of $\el$,
$$x^{\el-1}(1-x) - \frac{1}{k+\el} \leq \frac{1 }{c^{log_c(2k) + 1}}
\frac{1}{e} - \frac{1}{k + log_c(2k) + 2} $$
which can be seen to be non-positive for all $k$. This proves the following lemma.

\begin{lemma}
\label{lem:1item-ub}
Given any $D$ and $k$, let $\el = log_{\frac{e}{e-1}}(2k) + 2 \simeq \frac{\log{k}}{\log{\frac{e}{e-1}}} +
3.5$. Then,
$$\gain_D - \loss_D \geq 0$$
\end{lemma}

\vspace{1ex}
Thus, we have established the following theorem.
\begin{theorem}[One-item Upper Bound]
\label{thm:1item-ub}
  \ For the case of selling one item, we have
$$\eff(\rma(k+\el)) \geq \eff(\ema(k))$$ for any $k$ and $\el \geq
\left\lfloor\log_{\frac{1}{\alpha}}{2k} \right\rfloor + 2$, where
$\alpha = 1- 1/e$. Thus, $O(\log{k})$ extra bidders suffice for the
revenue maximizing mechanism to achieve at least as much efficiency
as the efficiency maximizing mechanism with $k$ bidders.
\end{theorem}

\subsection{Lower bound on the number of extra bidders required}
\label{subsec:1-item-lb} In this section, we will prove a lower
bound on the number of extra bidders $\el$ needed for the revenue
maximizing mechanism to achieve at least as much efficiency as the
efficiency maximizing mechanism with $k$ bidders.  To prove such a
lower bound, it suffices to specify a distribution $D$ s.t. the
contribution of the $\el$ extra bidders to expected efficiency of
$\rma(k+\el)$ is no more than $\gain$ and show that $\gain - \loss <
0$ for this choice of $k$ and $\el$.

Consider the distribution ${\G}_{\tfstar,\rstar}$ for any choice of
$\rstar, \fstar$ (see definition~\ref{def:Gfr}), and arbitrarily
small $\epsilon$. We first show that for the distribution specified
by ${\G}_{\tfstar,\rstar}(x)$, the contribution of $\el$ extra
bidders to the efficiency of $\rma(k+\el)$ is arbitrarily close to
$\gain_{{\G}_{\tfstar,\rstar}}$. To see this, note that the maximum
value possible under distribution ${\G}_{\tfstar,\rstar}(x)$ is
$\rstar + \epsilon$; thus, the maximum possible efficiency for any
draw of bidder values is $\rstar + \epsilon$. 
When all the first $k$ bidders draw a value below $\rstar$, the
$\el$ extra bidders contribute a maximum of $\rstar + \epsilon$ to
the efficiency of $\rma(k+\el)$ with probability equal to
$(1-{\G}_{\tfstar,\rstar}^{\el})$. Thus, the total contribution of
the $\el$ extra bidders to the efficiency of $\rma(k+\el)$ is
arbitrarily close to $\rstar(1-{\G}_{\tfstar,\rstar}^{\el})$, which
is the same as $\gain$.

Let $\alpha = 1-1/e$. Let
$\el(k)=\left\lfloor\log_{1/\alpha}{(k+1)(1-\alpha)}\right\rfloor +
1$.
Next we'll show that $\gain_{{\G}_{\tfstar,\rstar}} -
\loss_{{\G}_{\tfstar,\rstar} }  < 0$ for some choice of $\fstar$, any $k$ and for all
$\el \leq \el(k)$. To do this, recall the polynomial $q(x) = x^{k+\el}+
\ln(1-x) + \sum_{i=1}^k \frac{x^i}{i}$. By
Equation~\ref{eqn:gain-loss-final}, we know that $\gain_{{\G}_{\tfstar,\rstar}} -
\loss_{{\G}_{\tfstar,\rstar} } = \frac{ - \rstar
  q(\fstar)}{\fstar^k}$. Therefore, we just need to show that $q(x) > 0$ for
the above choice of $\el$ and some $x$ (we will choose $\fstar$ to be that $x$).
\begin{align}
q(\alpha) & =   \alpha^{k+\el(k)}+ \ln(1-\alpha) + \sum_{i=1}^k
\frac{\alpha^i}{i} \notag\\
 &=  \alpha^{k+\el(k)} -  \sum_{i=1}^\infty \frac{\alpha^i}{i} + \sum_{i=1}^k\frac{\alpha^i}{i}\notag \\
 & = \alpha^{k+\el(k)} -  \sum_{i=k+1}^\infty \frac{\alpha^i}{i} \notag \\
 & > \alpha^{k+\el(k)} - \frac{1}{k+1} \sum_{i=k+1}^\infty \alpha^i \notag\\
 & = \alpha^{k+\el(k)} - \frac{\alpha^{k+1}}{(k+1)(1-\alpha)} \notag\\
 & \geq 0 \notag
\end{align}
where the last inequality follows from the choice of $\el$.

This proves the following theorem.

\begin{theorem}[One-item Lower Bound]
\label{thm:1item-lb} Let $\alpha = 1-1/e$ and let
$\el(k)=\left\lfloor\log_{1/\alpha}{(k+1)(1-\alpha)}\right\rfloor +
1$. Then, if bidders are drawn i.i.d. from the distribution
described by $G_{\alpha, \rstar}$, $\eff(\rma(k+\el)) <
\eff(\ema(k))$ for any $r$, any $k$ and any $\el \leq \el(k)$. In
other words, $\el(k)$ extra bidders do not suffice for the revenue
maximizing mechanism to achieve as much efficiency as the efficiency
maximizing mechanism with $k$ bidders.
\end{theorem}

Note that $\el(k) \simeq
\frac{\log{k+1}}{\log{\frac{e}{e-1}}} - 2.2$, which differs from the
upper bound only by a small additive constant, namely $5.7$.

\section{Extension to multiple items}
\label{sec:multiitem}
In this section we consider the case of selling $t$ identical
items. As before, the efficiency maximizing mechanism gets $k$
bidders, and we want to find the smallest number of extra bidders that
suffice for the revenue maximizing auction to make as much
efficiency. As seen in the previous section, $\el = \el(k)=
\left\lfloor\log_{\frac{1}{\alpha}}{2k} \right\rfloor + 2$
suffice when selling only one item. So clearly, $\el t$ extra bidders
would suffice in the case of $t$ identical items. However, as we show
below, it is possible to prove a much tighter bound.

The question is to find the smallest number $s = s(k)$ such that the
efficiency of the revenue maximizing mechanism with $k+\el+s$
bidders, $\rma(k+\el+s)$, is at least as much as the efficiency of
the efficiency maximizing mechanism on $k$ bidders, $\ema(k)$, where
$\el = \el(k)$. As before, to compare the two mechanisms with
different number of bidders, we will think of the process of drawing
the values of $k+\el+s$ bidders $(b_1,b_2,..b_{k+\el+s})$
independently from the given distribution $D$; the first $k$ bidders
($b_1, b_2,..b_k$) participate in both $\rma(k+\el+s)$ and
$\ema(k)$, whereas the last $\el+s$ bidders participate only in
$\rma(k+\el+s)$.

We partition the space of draws of bidder values into $k+1$ parts: For
$i = 0,...,k$, the $i$th part, $\Omega_i$, consists of those draws in
which exactly $i$ of the first $k$ bidders have values greater than
$\rstar_D$. 
We now focus on one of these parts, say, the $i$th part $\Omega_i$,
and try to determine the expected efficiency of $\ema(k)$ and
$\rma(k+\el+s)$ over this restricted space. Wlog, we may assume that $i
< t$ (if $i \geq t$, then the efficiency of $\rma(k)$ is already equal
to that of $\ema(k)$). We define $t' = t-i$ and $k' = k-i$.

Let $(b_{max_1}, b_{max_2},\cdots, b_{max_k})$ be the bids of first
k bidders in the decreasing order of their value. Also let
$\common_i$ denote the sum of the highest $i$ of these bids,
conditioned on being in $\Omega_i$.

The expected efficiency of $\ema(k)$ conditioned on being in
$\Omega_i$, $\eff(\ema^i(k))$, equals
\begin{align}
&= E[ \sum_{j=1}^i b_{max_j}~|~ \Omega_i] + E[\sum_{j=i+1}^k b_{max_j}~|~ \Omega_i] \notag\\
&\leq \common_i + t' E[b_{max_{i+1}} ~|~ \Omega_i] \notag
\end{align}

Now,
\begin{align}
&E[b_{max_{i+1}} ~|~ \Omega_i] = \notag\\
&E[max(b_{i+1}, b_{i+2},\cdots, b_k) ~|~ b_1..b_i \geq r ~\&~
b_{i+1}..b_k < r] \notag\\
&~~~~~~~~~~~~\text{(because of the symmetry)} \notag\\
&=E[max(b_{i+1}, b_{i+2},\cdots, b_k) ~|~ ~
b_{i+1}..b_k < r] \notag\\
&~~~~~~~~~~~~ \text{(since $b_j's$ are independent)} \notag\\
&= \frac{\int_0^{\rstar} x f^{(k')}(x)dx}{F^{k'}(r)} \notag
\end{align}

Therefore,
\begin{align}
\label{eq:multi-EMA-bound}
&\eff(\ema^i(k)) \leq \common_i +
t'\frac{\int_0^{\rstar} x f^{(k')}(x)dx}{F^{k'}(r)}
\end{align}

Conditioned on being in $\Omega_i$, the contribution to the expected
efficiency of $\rma(k+\el+s)$ from the first $k$ bidders is
$\common_i$. The contribution to the expected efficiency of
$\rma(k+\el+s)$ from the remaining $\el+s$ bidders depends on how many of
the extra bidders have value above $\rstar$. If $j$ of these bidders
have value more than $\rstar$, then the contribution is at least
$\min(j,t') \rstar$. We have the following lower bound on this
contribution (note that the contribution from the extra $\el+s$ bidders
is independent of the first $k$ bidders, and hence of $\Omega_i$):

\begin{lemma}
\label{lem:l+s}
For any $\epsilon >0$, and for large enough $\el$, if $s
  \geq \left(t + (1+\epsilon)t\log{\el}\right)$, then the total
  contribution to the efficiency of $\rma(k+\el+s)$ from the remaining
  $\el+s$ bidders is at least $\rstar t' (1 - F^{\el}(\rstar))$.
\end{lemma}
\begin{proof}
  Recall that $\fstar = F(\rstar)$. Let $a_j = {\el+s \choose j}
  \fstar^{\el+s-j} (1-\fstar)^j$.  The contribution is:
  \begin{align}
    &  \rstar \left[ \sum_{j=0}^{t'-1} j a_j +
      (1 - \sum_{j=1}^{t'-1} a_j) t' \right]\notag\\
    = ~ &\rstar \left[ t' - \sum_{j=0}^{t'-1} a_j (t'-j) \right] \notag
  \end{align}
  But,
  \begin{align}
    a_j (t'-j) & \leq \frac{(\el+s)^j }{{j}!} \fstar^{\el} \fstar^{s-j}(1-\fstar)^j (t'-j)\notag\\
    & \leq \fstar^{\el} (\el+s)^j (1-1/e)^{s} (t'-j) \notag\\
    & \leq \fstar^{\el} \notag
  \end{align}
  for $s \geq \left(t + (1+\epsilon)t\log{\el}\right) + \log{t}$ and any
  $\epsilon>0$ and large enough $\el$. Thus the
  contribution is at least $\rstar t' (1 - \fstar^{\el})$.
\end{proof}
\vspace{1ex}

Let $\el =log_{e/(e-1)}(2k) + 2 $. Then, using Lemma~\ref{lem:l+s} and
the discussion above, the
efficiency of $\rma$ conditioned on being in $\Omega_i$ is:
\begin{align}
\eff(\rma^i(k+\el+s)) & \geq \common_i + t' (1-F(\rstar)^{\el})\rstar \notag\\
& \geq \common_i + t' \frac{\int_0^{\rstar} x f^{(k')}(x)
  dx}{F^{k'}(\rstar)} \notag \\
  &\mbox{(by Lemma~\ref{lem:1item-ub}, and since $k' \leq k$)}\notag\\
& \geq \eff(\ema^i(k))\notag \\
&~~~~~~ \mbox{(by equation (\ref{eq:multi-EMA-bound}))}\notag
\end{align}

Thus we have proved the following theorem (we have not tried to
optimize the constants or how large $k$ has to be for this result to
hold).
\begin{theorem}[Multi-item Upper Bound]
\label{thm:multiitem-ub}
In the case of selling $t$ identical items, we have
$$\eff(\rma(k+\el+s)) \geq \eff(\ema(k))$$
for $\el\geq \left\lfloor\log_{\frac{1}{\alpha}}{2k} \right\rfloor + 2$
and $s \geq \left(t + (1+\epsilon)t\log{\el}\right)$, for every $\epsilon
>0$ and large enough $k$. Thus,
approximately, $\log{k} + t \log\log{k}$ extra bidders suffice.
\end{theorem}

\section{The case of regular distribution}
\label{sec:regular}
In this section, we will show that for any given $k$ and $\el$,
there exists a regular distribution $D$ s.t. expected efficiency of
$\rma(k+\el)$ is less than the expected efficiency of $\ema(k)$.

To recall, a distribution $D$ is regular if and only if the function
$\psi(x):= x - \frac{1}{h_D(x)}$ is non decreasing in $x$. Now
consider the following distribution:

\begin{equation*}
  P_{\epsilon, r}(x)   :=
 \begin{cases}   1 - \frac{\epsilon}{x+\epsilon} & \text{~~~~$0 \leq x < r$,}
     \\
     1 &\text{~~~~$x \geq r$}
  \end{cases}\notag
\end{equation*}

One can easily verify that the above distribution is regular for
every choice of $\epsilon,r > 0$, by evaluating its $\psi$ function.
Moreover the reserve price of this distribution equals $r$. Now,
using similar arguments as used in the previous sections, we can
show that the contribution of the extra $\el$ bidders to
$\eff[\rma(k+\el)]$ is $r(1 - (1- \frac{\epsilon}{r+\epsilon}
)^{\el})(1- \frac{\epsilon}{r+\epsilon} )^{k} ~\leq~ r(1 - (1-
\frac{\epsilon}{r+\epsilon} )^{\el}) $. Also, the extra contribution
of the first $k$ bidders to $\eff[\ema(k)]$ over $\eff[\rma(k+\el)]$
when all of the first $k$ bidders have a value below reserve price
$r$ is $ \int_0^r x [P_{\epsilon, r}^k(x)]' dx$.

Now, as we decrease the value $\epsilon$, the term $r(1 - (1-
\frac{\epsilon}{r+\epsilon} )^{\el})$ decreases and $ \int_0^r x
[P_{\epsilon, r}^k(x)]' dx$ increases for a fixed $k$ and $\el$.
Moreover, one can show that there exists a small enough $\epsilon :=
\epsilon'$ such that $ \int_0^r x [P_{\epsilon', r}^k(x)]' dx$ is more
than $r(1 - (1- \frac{\epsilon'}{r+\epsilon'} )^{\el})$. Thus, for the
distribution $P_{\epsilon', r}$ , the loss in $\eff[\rma(k+\el)]$
because of the reserve price is more than the gain from extra $\el$
bidders.\\

\noindent{\bf Acknowledgment:} We thank Hal Varian for useful
comments and discussions. \\
%

\end{document}